\documentclass[twocolumn]{aastex631}

\newcommand{\snrem}{0509$-$67.5}
\newcommand{\snremspace}{0509$-$67.5 }

\shorttitle{SNR \snremspace Surviving Companion Search}
\shortauthors{Shields et al.}
\graphicspath{{./}{figures/}}

\usepackage{graphicx}
\usepackage{subfigure}
\usepackage{amsmath}

\usepackage{savesym}
\savesymbol{tablenum}
\usepackage{siunitx}
\restoresymbol{SIX}{tablenum}

\begin{document}

\title{No Surviving SN Ia Companion In SNR \snrem: Stellar Population Characterization and Comparison To Models}

\author[0000-0002-1560-5286]{Joshua V. Shields}
\affiliation{Department of Physics and Astronomy, Michigan State University, East Lansing, MI 48824, USA}
\email{shield90@msu.edu}

\author[0000-0002-6688-3307]{Prasiddha Arunachalam}
\affiliation{Department of Physics and Astronomy, Rutgers University, 136 Frelinghuysen Road, Piscataway, NJ 08854, USA}

\author[0000-0002-0479-7235]{Wolfgang Kerzendorf}
\affiliation{Department of Physics and Astronomy, Michigan State University, East Lansing, MI 48824, USA}
\affiliation{Department of Computational Mathematics, Science, and Engineering, Michigan State University, East Lansing, MI 48824, USA}

\author[0000-0002-8816-6800]{John P. Hughes}
\affiliation{Department of Physics and Astronomy, Rutgers University, 136 Frelinghuysen Road, Piscataway, NJ 08854, USA}

\author[0009-0004-2531-7423]{Sofia Biriouk}
\affiliation{Department of Physics and Astronomy, Michigan State University, East Lansing, MI 48824, USA}

\author[0009-0009-6406-6942]{Hayden Monk}
\affiliation{Department of Physics and Astronomy, Michigan State University, East Lansing, MI 48824, USA}

\author[0000-0003-0426-6634]{Johannes Buchner}
\affiliation{Max-Planck-Institut fur extraterrestrische Physik, 
Giessenbachstrasse 1, 85748 Garching bei Munchen, Germany}



\begin{abstract}

The community agrees that Type Ia supernovae arise from Carbon/Oxygen white dwarfs undergoing thermonuclear runaway. However, the full progenitor system and the process that prompts the white dwarf to explode remain unknown. Most current models suggest that the white dwarf explodes because of interaction with a binary companion which may survive the process and remain within the resulting remnant of the exploded star. Furthermore, both the pre-supernova interaction process and the explosion of the primary are expected to imprint a significant departure from ordinary stellar radii and temperatures onto the secondary, making the star identifiable against the unrelated stellar population. Identification of a surviving companion inside an SN Ia remnant might confirm a specific corresponding SN Ia progenitor channel based on the identity of the companion. We conducted a surviving companion search of the Type Ia remnant SNR \snrem based in the Large Magellanic Cloud. The well-constrained distance to and foreground extinction of the Large Magellanic Cloud allow for Bayesian inference of stellar parameters with low correlation and uncertainties. We present a deep catalog of fully characterized stars interior to SNR \snremspace with radii, effective temperatures, and metallicities inferred using combined Hubble Space Telescope photometric observations across multiple visits.
We then compile a list of surviving companion models appropriate for the age of the remnant (roughly 400 years after the explosion). We compare these predictions with the inferred stellar parameters and conclude that none of the stars are consistent with the predicted signatures of a surviving companion.

\end{abstract}


\section{Introduction} \label{sec:intro}

Type Ia supernovae (SNe Ia) play a critical role across astronomy. As standardizable candles, they serve as distance indicators \citep[e.g.][]{phillips_absolute_1993, phillips_reddening-free_1999} that led to the discovery of the accelerating expansion of the universe \citep{riess_observational_1998, schmidt_high-z_1998, perlmutter_measurements_1999}. They are also known to provide a substantial fraction of the iron group elements in the Universe, driving galactic chemical enrichment over cosmic time \citep{timmes_galactic_1995, nomoto_nucleosynthesis_2013, kobayashi_new_2020}. Despite the importance of these energetic events, we have still not confirmed their progenitor system and explosion mechanism \citep{livio_progenitors_2000, wang_progenitors_2012, ruiz-lapuente_surviving_2019}. It is widely accepted that SNe Ia arises from the thermonuclear explosions of Carbon/Oxygen (C/O) white dwarfs \citep{pankey_possible_1962, colgate_early_1969}, but the process that prompts the white dwarfs to explode is still uncertain. This major uncertainty on the origins of SNe Ia propagates directly to their empirically calibrated interpretations, introducing additional uncertainty or potential bias into our understandings of galactic chemical enrichment and universal expansion.

SN Ia progenitor models can be divided into two major channels distinguished by the survival or destruction of the secondary in the binary progenitor system that prompts the C/O white dwarf to explode. 
In the first channel, the primary white dwarf violently merges with a secondary white dwarf, prompting thermonuclear explosion \citep[e.g.][]{webbink_double_1984}. In this scenario, the explosion may fully unbind both stars leaving no stellar remnant behind \citep{pakmor_normal_2012, papish_response_2015}. In the second, the primary either stably or unstably accretes material from a secondary which prompts thermonuclear runaway in one of a variety of ways \citep{whelan_binaries_1973, iben_supernovae_1984}. 
The companion in this broad progenitor channel has been proposed to be either a helium star \citep{iben_helium-accreting_1994, wang_helium_2009}, a helium-rich white dwarf \citep{bildsten_faint_2007}, a post-main sequence giant star \citep{li_evolution_1997}, an evolved sub-dwarf \citep{meng_subdwarf_2019}, or a main sequence star \citep{han_companion_2008, meng_single-degenerate_2009}. The majority of accretion scenarios predict that the secondary survives the SN explosion and will remain near the site of the explosion \citep{shappee_type_2013}. Crucially, this means that the surviving companion to the exploded star will reside near the center of the supernova remnant (SNR) that is produced by the event. 

In most accretion scenarios, the accretion and explosion processes are expected to impart identifiable signatures onto the companion that persist for at least thousands of years: During mass transfer, the secondary may experience tidal forces that may both heat the star \citep{shen_wait_2017}. This process may also cause the secondary to become tidally locked with the primary which will also cause the star to adopt an anomalously high rotation rate \citep{kerzendorf_subaru_2009}. When the primary explodes, the secondary will no longer be gravitationally bound and will be shot out of the system at its pre-explosion orbital velocity \citep{shen_three_2018}, far above velocities obtainable through normal processes of stellar evolution. Shortly after the primary explodes, the companion will be impacted by the SN ejecta, which may strip a significant amount of material from the star \citep{marietta_type_2000}, or deposit energy into the stellar envelope, causing it to increase in temperature or inflate \citep{liu_observational_2021}. Any ejecta that is captured by the companion will be rich in radioactive elements which can continue to inject energy into the star \citep{shen_wait_2017}. The ejecta will also contain a large fraction of iron group elements which can remain detectable in the stellar photosphere \citep{ozaki_method_2006}. 


If a surviving companion can be identified inside an SNR Ia by a combination of the signatures expected for such a star, that identification would directly confirm a specific corresponding accretion scenario as a viable SN Ia progenitor channel \citep{marietta_type_2000, pakmor_impact_2008}. However, decades of intense, focused searches of Galactic SNRs have failed to securely identify even a single unambiguous SN Ia companion \citep[see e.g.][]{ruiz-lapuente_binary_2004, ruiz-lapuente_no_2018, ruiz-lapuente_surviving_2019, kerzendorf_high-resolution_2013, kerzendorf_search_2018, shields_searching_2022}. Unfortunately, it is difficult to extrapolate from these results to make conclusive statements about the viability of SN Ia progenitor channels. Galactic remnants are plagued with highly uncertain distances \citep[see e.g. Tycho's SNR,][]{ihara_searching_2007} and unconstrained column densities of foreground dust \citep[see e.g. Kepler's SNR,][]{kerzendorf_reconnaissance_2014} which leads to generally narrow searches that are not sensitive to all viable progenitor channels and the surviving companions they are expected to produce. 

In direct contrast, the Large Magellanic Cloud (LMC) resides at a well-constrained distance of $49.59 \pm 0.63$\,kpc \citep{pietrzynski_distance_2019}, with very little foreground extinction \citep{joshi_reddening_2019} due to its high Galactic latitude and face-on orientation \citep{van_der_marel_new_2002}. Additionally, the galaxy resides close enough for individual stars to be resolved by the Hubble Space Telescope (HST). These properties make SNRs Ia enclosed in the LMC uniquely well-suited for surviving companion searches that are free from the major problems of remnants in the Milky Way \citep{li_search_2019}. 

SNR \snremspace is a young LMC remnant confirmed to be of SN Ia origin by light echoes of the original SN spectrum \citep{rest_testing_2005, rest_spectral_2008}. \cite{schaefer_absence_2012} conducted a tight search near the geometric center of the remnant, with a search region large enough to discover stars ejected from the progenitor system at speeds of up to 390\,km\,s$^{-1}$. That work found no objects in their search region other than a diffuse source that was later identified to be a background galaxy \citep{litke_nature_2017}, suggesting that no surviving companion exists within the remnant. However, recent works allow for surviving companions traveling significantly faster \citep[see e.g.][up to 2500\,km\,s$^{-1}$]{shen_three_2018}, suggesting that a companion has not been conclusively ruled out. Recently, \citet[][hereafter A22]{arunachalam_hydro-based_2022} used accurate proper motion and location measurements of the remnant's forward shock to determine a precise dynamical explosion center. We decided to revisit this remnant with these new constraints to conduct a thorough surviving companion search. 

\begin{table*}[t!]

    \centering
    \begin{tabular}{c|c|c|c|c|c}
         Prop. ID & MJD &  Prop. PI & Filter & Exp. Time (s) & Limiting Mag (AB)\\
          \hline
         12326 & 55504 & Noll & F475W & 1010 & 26.08\\
         12326 & 55504 & Noll & F555W & 696 & 26.05\\
         12326 & 55504 & Noll & F814W & 800 & 25.17\\
         13282 & 56559 & Chu & F814W & 1465 & 25.49\\
         13282 & 56559 & Chu & F110W & 298 & 25.53\\
         13282 & 56559 & Chu & F160W & 798 & 25.03 \\
    \end{tabular}
    \caption{The six observations combined to create our stellar catalogs. Limiting magnitudes were estimated based on the faintest star observed in the observation.}
    \label{tab:obs}
\end{table*}

In this work, we present a characterized catalog of the stellar population inside SNR \snremspace to discover a surviving Ia companion. We synthesized six HST observations of the remnant in five broad filters to construct detailed spectral energy distributions (SEDs) of the enclosed stars. We fit the observations of each star with backward physical modeling using Bayesian Inference to move from SEDs to distributions of stellar astrophysical parameters that could produce each set of photometric observations. 

In Section \ref{sec:obs}, we describe the data and synthesis methods we used to construct SEDs of the interior stellar population to SNR \snrem. In Section \ref{sec:methods}, we describe our models and astrophysical parameter inference methods. In Section \ref{sec:results}, we present the results of our inference and compare them to the expected parameters of surviving companions in the literature. In Section \ref{sec:discussion}, we discuss the constraints our results place on allowed surviving companions in the SNR \snremspace remnant. We conclude in Section \ref{sec:conclusions} by highlighting the parameter space ruled out for surviving companions as well as discussing further implementations of the techniques in this paper.

\section{Observations} \label{sec:obs}

\begin{figure}[t!]
    \centering
    \includegraphics[width=.47\textwidth]{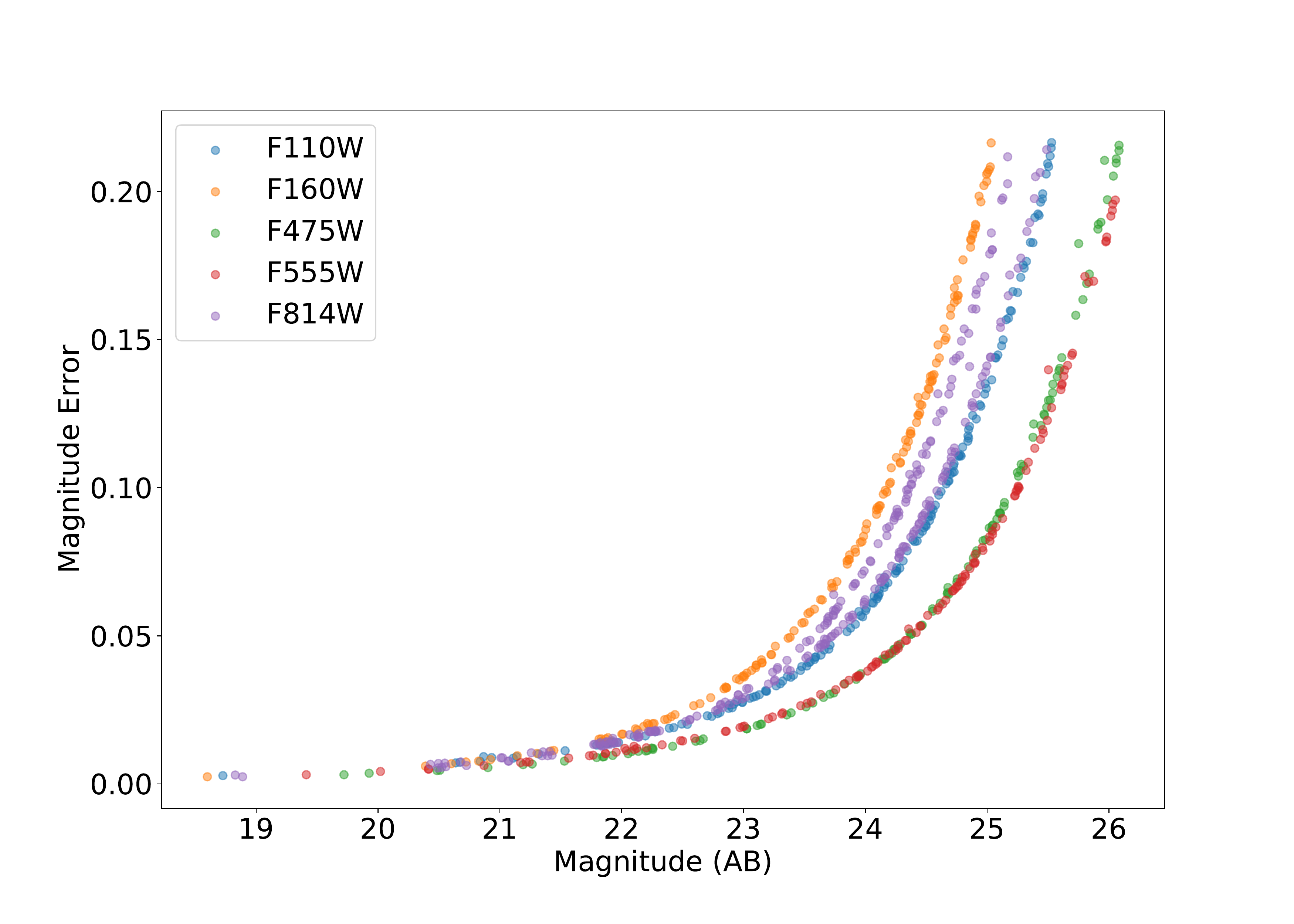}
    \caption{Error on magnitude as a function of magnitude for the full set of characterized stars in the larger sample, by filter.  These data show the observational uncertainties used for our Bayesian inference fitting. 
}
    \label{fig:errors}
\end{figure}

We needed to construct a well-sampled SED of each star inside SNR \snremspace to characterize the interior stellar population and search for the existence of a surviving companion. To measure stellar radii and temperatures, we fundamentally needed to constrain the peak and shape of each stellar blackbody curve. SNR \snremspace has been well observed by the HST in eight different filters. In this work, we chose to restrict ourselves to five wide-band HST filters (see Table \ref{tab:obs}) for ease of methodological development. A larger set of filters would more powerfully constrain stellar astrophysical parameters but is beyond the scope of this work. 

\subsection{HST Catalogs}

We set out to create a catalog of the interior stellar population to the SNR \snremspace using a large enough sample of HST observations to construct well-sampled SEDs. However, individual archival HST observations have been shown to suffer from misaligned WCS with errors on the order of an arcsec, which has previously prohibited the creation of stellar catalogs with straightforward coordinate-based source matching. 
However, the recent HST Astrometry Project has re-derived WCS for archival observations by cross-identifying bright stars Gaia DR2 for alignment and reduced astrometric errors to approximately 10 mas \citep{hoffmann_new_2021}, published in the Hubble Source Catalog (HSC). We queried updated stellar sky positions and matched sources in the HSC available through the \textbf{CasJobs} database \citep{whitmore_version_2016} for cross-visit matching to construct SEDs for stellar characterization. The resulting individual, unmatched observations are shown in Figure \ref{fig:errors}.

\subsection{Search Region}

We obtained observations of each star within a 4.2 arcsec (1.00\,pc at 49.27\,kpc) radius of the dynamical center of the SNR \snremspace as determined by A22. This radius corresponds to the distance that a 2500\,km\,s$^{-1}$ object (the fastest surviving companion velocity predicted in our considered models) could have traveled at the distance of the LMC, over a highly conservative age estimate for the remnant of 400 years  (a strong upper limit from models of the remnant's expansion, see A22 for further discussion). 

In addition, we obtained observations of the surrounding stellar population within 10 arcsec of the dynamical center of SNR \snremspace in order to have a control sample of stars that could not be the surviving companion. We used this sample to better understand the sensitivity of our search and how the individual limiting magnitudes of the observations translate to limits on inferred astrophysical parameters. 

\begin{figure*}[t]
    \centering
    \includegraphics[width=\textwidth]{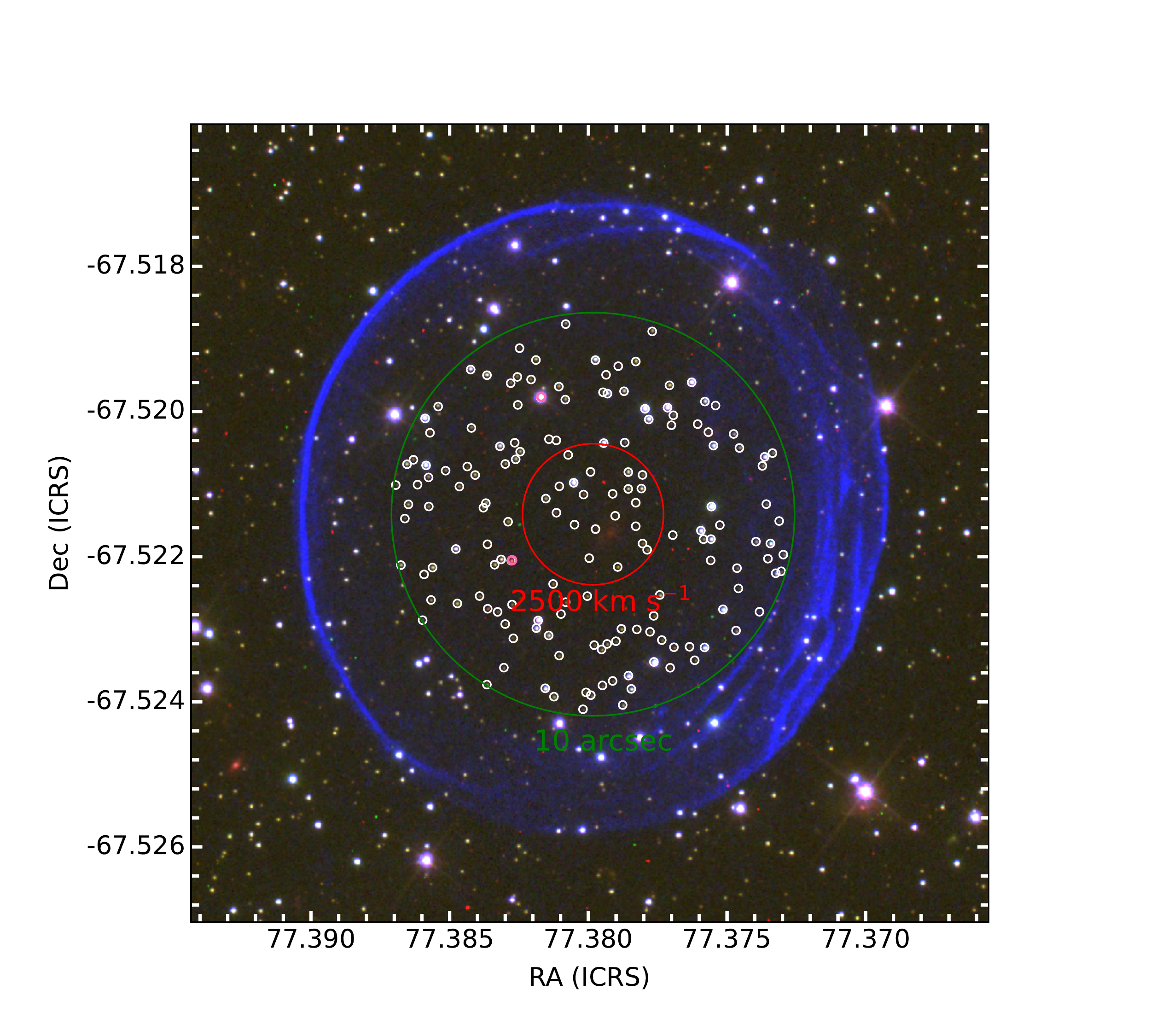}
    \caption{A composite image of SNR \snremspace from archival photometry. The red channel is the F814W filter, green is F555W, and blue is F656N. The red circle shows our search region where stars are close enough to the center to be the surviving companion, and the green region shows the larger control sample of characterized stars we used to better understand the unrelated stellar population and the uncertainties of our technique. The white circles show the individual stars that were modeled and characterized. The two large stars we discuss in Subsection \ref{subsec:large_rads} are circled in pink.  
}
    \label{fig:rem_viz}
\end{figure*}

In Figure \ref{fig:rem_viz}, we show a composite image of SNR \snremspace created with \texttt{APLPY} \citep{robitaille_aplpy_2012} using two of the six observations analyzed in this work (F814W and F555W), as well as F656N to clearly show the remnant. We show the search region of potential surviving companions, the control sample, as well as each of the identified interior stars. 

\section{Methodology \& Analysis} \label{sec:methods}

Our goal was to be sensitive to the entire range of predicted surviving companions which we gathered and show in Table \ref{tab:models}. In designing this study, we chose to propagate the most conservative case for each parameter across all other parameters to be certain that we were sensitive to any potential type of surviving companion. As a result, certain portions of our combined parameter space are not realizable for a surviving companion (e.g. a non-degenerate companion cannot reside far away from the center of the remnant due to a lower maximum ejection speed). However, we decided that this approach was necessary to safeguard against potentially missing a surviving companion.

\subsection{Stars Identified in A22}

We note the presence of three stars identified in A22 as potential surviving companion candidates that reside very near the dynamical center of SNR \snrem, within 1.4 arcsec, the angular distance that an object in the LMC moving at 1000\,km\,s$^{-1}$ could have traveled in 317 years. These three stars are too faint to be detected by the automated aperture photometry that created the HSC. For completeness, we decided to include these stars in our search using HST magnitudes from \cite{hovey_kinematic_2016}, but we note that their photometry was obtained inconsistently with the rest of the stellar population. We discuss these stars further in Subsection \ref{subsec:faint_stars}.

\subsection{Synthetic Spectra and Photometry}

\begin{table*}[t]

    \centering
    \begin{tabular}{c|c|c|c|c}
         Source Work & Model & Surviving Companion Type & Temperature ($10^{4}$ K) & Radius (R$_\odot)$\\
          \hline
         \citealt{liu_observational_2021} & DDet$^*$ Kepler Model & He Star & 1 & 3\\
          \citealt{liu_observational_2021} & DDet$^*$ Mesa Model & He Star & 1.1 & 0.5\\
          \citealt{liu_signatures_2022} & SD Accretion$^\ddagger$ & He Star & 4.6 & 0.5\\
          \citealt{shappee_type_2013} & SD Accretion$^\ddagger$ & MS Star & 0.52 & 10\\
          \citealt{pan_evolution_2013} & SD Accretion$^\ddagger$ & He Star & 4 & 0.6\\
          \citealt{pan_search_2014} & SD Accretion$^\ddagger$ & MS Star & 0.75 & 3.42\\
          \citealt{pan_search_2014} & DDet$^*$ & He WD & 5.5 & 0.39\\
          \citealt{rau_evolution_2022} & SD Accretion$^\ddagger$ & MS Star & 0.4 & 2.5\\
          \citealt{shen_wait_2017} & DDet$^*$ & He WD & 10 & 0.05$^{\dag}$\\

    \end{tabular}
    \caption{
    A summary of predicted temperatures and radii from different surviving companion models at the age of the remnant $\leq 1000$\,yrs. Many of these works tested a variety of models. In such cases, we show the resulting temperature and radius that is the most difficult to detect, i.e. the smallest, coolest model. We show two models from works where the smallest model is not also the coolest. All combinations of temperature and radius are clearly distinguishable from the surviving companion candidates inside SNR \snrem with the exception of \cite{shen_wait_2017} which is too faint to detect. \\
    $^*$ DDet refers to the Double Detonation model, also known as the Dynamically Driven Double-Degenerate Double-Detonation (D6) model \citep{shen_three_2018}.\\
    $^\ddagger$ SD Accretion refers to the single degenerate accretion scenario, in which the companion has exhausted fusion energy.\\
    $^{\dag}$\cite{shen_wait_2017} did not directly present an evolving radius of the surviving companion over time. We calculated a radius using the provided luminosity and effective temperature using the Stefan-Boltzmann law. 
    }
    \label{tab:models}
\end{table*}

We generated synthetic spectra by interpolating a spectral grid generated with the \texttt{PHOENIX} stellar atmosphere code \citep{husser_new_2013} using \texttt{StarKit}\footnote{\url{https://github.com/starkit/starkit}} \citep{kerzendorf_starkit_2015}. The \texttt{PHOENIX} spectra to be interpolated were sampled over effective temperature {$T_{\textrm{eff}}$}, logarithmic surface gravity log$(g)$, and metallicity relative to solar $[\frac{M}{H}]$. We then scaled the spectrum by stellar radius as a free parameter to obtain a model intrinsic stellar spectrum. 

We attributed all extinction to Galactic foreground dust, which is the dominant source of extinction for objects contained in the LMC \citep{choi_smashing_2018}. We reddened our stellar spectrum with a standard Galactic extinction law of $R_V =3.1$ following \cite{fitzpatrick_correcting_1999} with the \texttt{dust\_extinction}\footnote{\url{https://github.com/karllark/dust\_extinction}} package. We note that, because of the modest cumulative amount of foreground extinction (A$_V \approx 0.28$, \citealt{joshi_reddening_2019}), the exact choice of reddening model had little effect on synthetic magnitudes generated and resulting extracted stellar parameters (roughly a maximum difference of 0.01 mags in F475W and F814W, with smaller effects in other bands).
Finally, we convolved the spectrum with each HST filter curve in which a given star was observed (shown in Figure \ref{fig:fit_example}) using \texttt{wsynphot}\footnote{\url{https://github.com/starkit/wsynphot}} \citep{kerzendorf_wsynphot_2022}, and integrated the resulting flux to obtain a photometric magnitude that we compared directly to each observation.

\subsection{Bayesian Parameter Inference}
\label{subsec:bayes}

 We derived a multidimensional posterior probability distribution for the set of astrophysical parameters for each star with the nested sampling Monte Carlo algorithm MLFriends \citep{buchner_statistical_2016, buchner_collaborative_2018} implemented in the \texttt{UltraNest}\footnote{\url{https://johannesbuchner.github.io/UltraNest/}} package \citep{buchner_ultranest_2021} by comparing the observations to synthetic magnitudes generated with a given set of parameters. Our priors are presented in Table \ref{tab:priors}. We fixed log$(g)$ at 3.0 as our observations were generally insensitive to variation in this space. Additionally, we fixed each star to be at a distance of 49.27\,kpc, the distance of the remnant following A22, noting that any variation in distance for a given star in the LMC is  small compared to other sources of uncertainty, and any variation or uncertainty here translates directly to a corresponding variation or uncertainty in radius. An uncertainty on distance of 1\,kpc (2\%) typical for the LMC, would translate directly to an uncertainty in radius of 2\%.

We begin with a standard $\chi^2$ likelihood assuming Gaussian distributed errors for our observations given by
\begin{equation*} 
\chi^2 = \sum_{i=1}^n \left(\frac{m_{i,obs} - m_{i,model}}{\sigma_{i,obs}}\right)^2 
\end{equation*}
over the observations for a given star. We adopt the log-likelihood function $ln\mathcal{L} = \chi^2/2$ for fitting.



\begin{figure}[t]
    \centering
    \includegraphics[width=.45\textwidth]{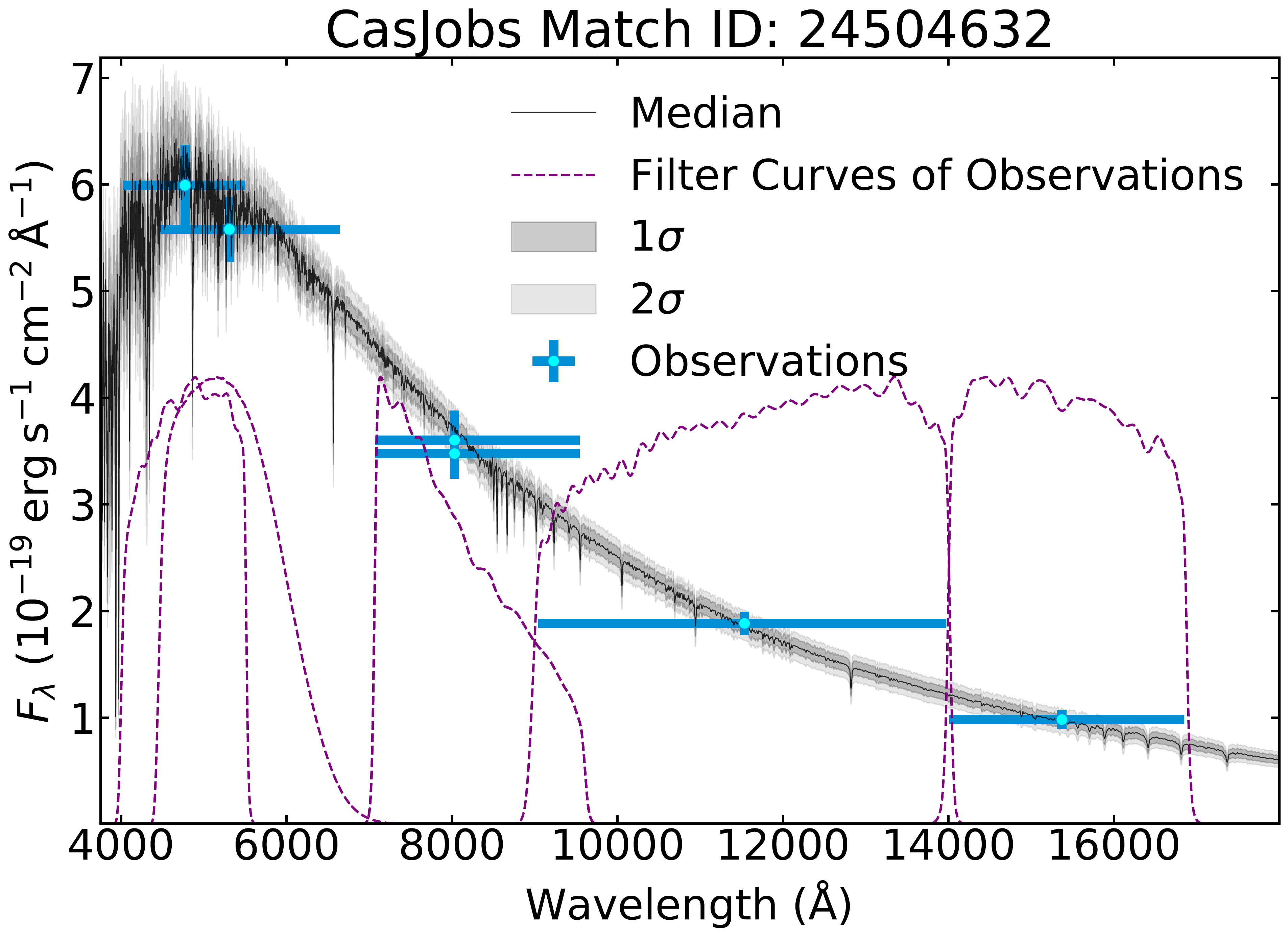}
    \caption{An example outcome of the fitting process for one of the stars in the search region. The blue crosses show the photometric observations with the vertical spread showing observational uncertainty, and the horizontal spread shows the width of the filter used for the observation, corresponding to the rough portion of the spectrum probed by the observation. The transmission curve of each filters is shown in dashed purple. The black and grey regions show a distribution of spectra created by sampling the posterior distribution of constrained stellar parameters.}
    \label{fig:fit_example}
\end{figure}

We show an example of a single set of observations and the resulting fit in Figure \ref{fig:fit_example}. The distribution of spectra in this figure is generated by randomly sampling the posterior parameter distribution and generating model spectra as examples to show the constraints of the input observations. 

\subsection{SED Requirements}
Stars only detected in the infrared F110W and F160W bands did not have sufficient observational constraints to allow for strong astrophysical characterization. As a result, we only present characterizations of stars with observations in at least two non-infrared bands. This generally required that stars be brighter than 25th magnitude, seen in the limiting magnitude estimation reported in Table \ref{tab:obs}.

\subsection{Validation, Sensitivity, and Precision of Parameter Extraction}

\begin{figure*}[t]
    \centering
    \includegraphics[width=\textwidth]{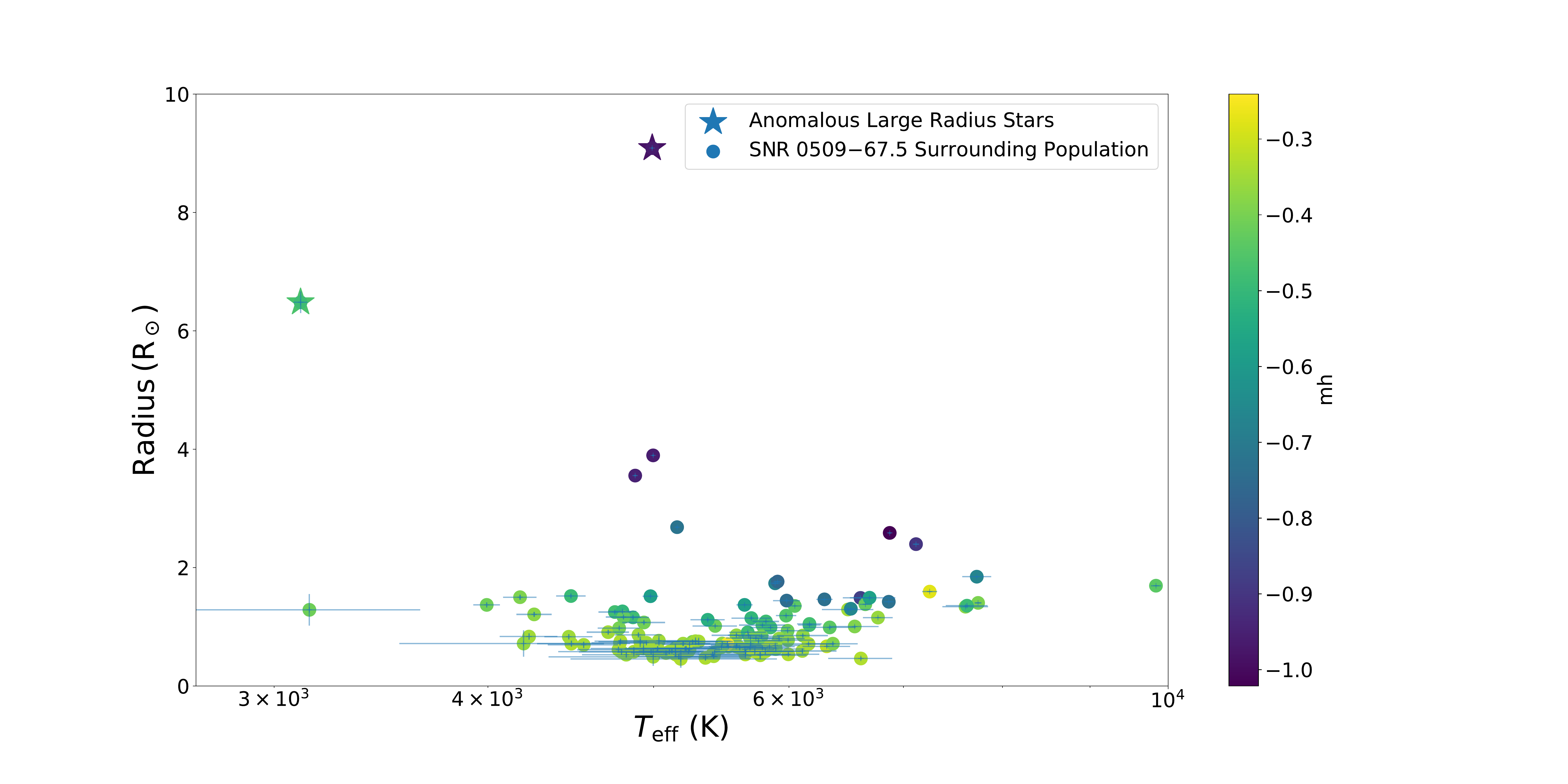}
    \caption{Radius vs. Effective Temperature and metallicity fits of stars in a 10 arcsec radius, excluding those close enough to the center of SNR \snremspace to be the surviving companion. We used this sample to estimate the underlying main sequence distribution of stars in the local LMC. We note the presence of two anomalous large radius stars that we do not support belonging to the local main sequence population, and thus do not include in our distribution estimation in figure \ref{fig:teff_rad}. We discuss these stars further in Subsection \ref{subsec:large_rads}.
}
    \label{fig:teff_rad_large}
\end{figure*}

\begin{figure*}[t]
    \centering
    \includegraphics[width=\textwidth]{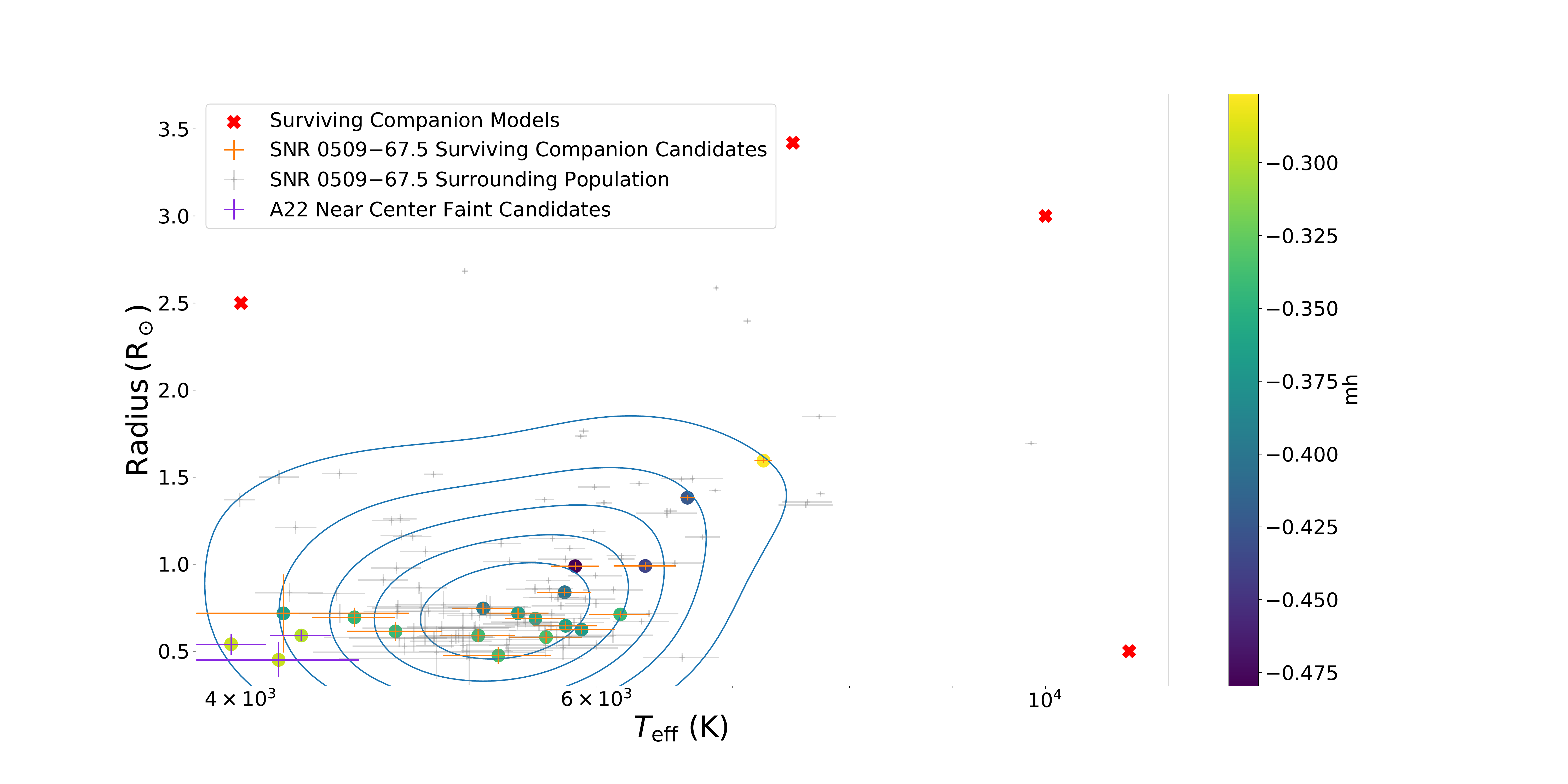}
    \caption{Radius vs. Effective Temperature and metallicity fits of stars within 4.2 arcsec of the hydrodynamic center of SNR \snrem, the sample of potential surviving companion stars. The gray crosses show the surrounding control sample from Figure \ref{fig:teff_rad_large} which we used to estimate the underlying temperature-radius distribution of the local stellar population, shown by the contours. The purple crosses show the three stars identified in A22. The red X markers show surviving companion models from Table \ref{tab:models} close enough to be compared, with the other models too high in temperature or large in radius to be placed within the bounds of the figure. 
}
    \label{fig:teff_rad}
\end{figure*}

We examine the constraining power of our parameter extraction and find robust validation of our ability to recover known temperatures and radii. In Appendix \ref{app:sun_val}, we fit the sun as an example and show the posterior parameter distribution. We find that the precision of our metallicity constraint is highly dependent on the temperature of the star, and in some cases virtually no information is gained and the prior is recovered. In principle, a star enriched by over an order of magnitude in metallicity, a potential signature of a surviving SN Ia companion, could be constrained and detected with HST photometry. However, no star in our sample showed an obvious metallic enhancement signature (See appendix \ref{app:metallicity_val} for further discussion). For this reason, we show the maximum likelihood metallicities that result from our fitting process in Figures \ref{fig:teff_rad_large} and \ref{fig:teff_rad}, but do not further probe this parameter space for the signature of a surviving companion. Similarly, while we fit for foreground extinction and thus allow the variation in the parameter to inform the posterior, in almost all cases we recovered our extinction prior. 

In this work, we did not attempt to characterize a systematic uncertainty floor for our parameter constraints. We note that the brightest stars in our sample show unrealistically small estimated temperature uncertainties, smaller than our models are truly able to constrain tracing back to uncertainty in our model \texttt{PHOENIX} spectra. We report these values as they were produced and note that they do not influence our search for a surviving companion, but suggest caution when interpreting these uncertainties for other purposes. 

Our various limiting magnitudes in the original observations do not easily map onto uncorrelated temperature and radius sensitivities. We estimate the sensitivity of our HSC search in the dependent parameter space as the smallest, coolest stars in our sample. For a 5000\,K star this results in a sensitivity to stars larger than half a solar radius, scaling with temperature. A 10000\,K star in line with the predictions of \cite{liu_signatures_2022} would need to be smaller than 0.2\,R$_\odot$ to remain undetected.

\section{Results} \label{sec:results}

\subsection{Characterized Stellar Astrophysical Parameters}

We modeled the interior stellar population of SNR \snremspace to extract stellar parameters and search for a surviving companion in line with those predicted by the surviving companion models shown in Table \ref{tab:models}. We extracted the parameters of each star within a 10 arcsec radius of the dynamical center, too far from the center of the remnant to be the surviving companion to the exploded star, to serve as a control sample of the local stellar population, shown in Figure \ref{fig:teff_rad_large}. We then analyzed and extracted the parameters of each star within a conservative 4.2 arcsec search region, the maximum distance a surviving companion could have traveled from the center of the remnant. This is the sample of viable surviving companion candidates, shown in Figure \ref{fig:teff_rad}. In this sample, we do not detect a star substantially different from the local temperature-radius main sequence or in line with any of the models shown in Table \ref{tab:models}. 

\begin{table*}[!t]
    \centering
    \begin{tabular}{c|c|c}
         Parameter & Prior Distribution & Source \\
          \hline
          Effective Temperature (K) &  $\mathcal{U}(2300-12000)$ & Phoenix Grid Boundaries\\
          Radius (R$_\odot$) & $\mathcal{U}(0.05-10)$ & Chosen Physical Boundaries\\
           V band Absorption (mags) & $\mathcal{N}(0.28, 0.15)$ bounded at 0 & \citealp{joshi_reddening_2019}\\
          Metallicity ($[\frac{M}{H}]$) & $\mathcal{N}(-0.34, 0.15)$ & \citealp{luck_magellanic_1998}\\
          Distance (kpc) & 49.27 & \citealp{arunachalam_hydro-based_2022}
          
    \end{tabular}
    \caption{A compilation of the priors used for stellar parameter inference, as well as sources where appropriate. $\mathcal{U}$ is Uniform (Lower boundary - Upper boundary), $\mathcal{N}$ is Normal($\mu, \sigma$), where $\mu$ is the mean and $\sigma$ is the standard deviation. See subsection \ref{subsec:bayes} for discussion of the choice of priors.}

    \label{tab:priors}
\end{table*}

\subsection{Large Radius Stars in the Control Sample}
\label{subsec:large_rads}

 We note the presence of two anomalously large stars in our greater control sample of 10 arcsec around the dynamical center of SNR \snremspace denoted separately in Figure \ref{fig:teff_rad_large}. An inflated star could be a tracer of a surviving companion, however neither of these stars lie close enough to the dynamical center of the remnant to be associated with the Type Ia event. We highlight that a large radius here is not necessarily physical because radius is degenerate with distance and all stars were assumed to be at the distance of the SNR \snremspace  for fitting. We favor the explanation that stars with anomalously large radii here are either foreground Galactic halo stars or red giants. 
 
 We examined the Besancon Galaxy model \citep{robin_besancmodel_2012} to investigate the possibility of observing a foreground Galactic halo star in our sample. We sampled 10 arcsec cones oriented towards the LMC and found a stellar surface number density of $1.1 \times 10^{-3}$ arcsec$^{-2}$, or 0.359 stars in a 10 arcsec radius circle on average. Assuming star counts are Poisson distributed, we calculate a $30.2\%$ chance to observe one or more Galactic halo stars, and a $5.1\%$ chance to observe 2 or more. Assuming the 3000\,K star in Figure \ref{fig:teff_rad_large} is in the foreground, scaling its radius appropriately by roughly a factor of 10 for a halo star would place it firmly on the main sequence, making this a likely explanation. Additionally, A22 identified this star (also known as star M in \citealt{schaefer_absence_2012}) as possessing the largest proper motion in and around the remnant, consistent with it residing in the foreground and further strengthening this identification. 
 
 The star that we measured to have a radius of 9.5\,R$_\odot$ and a temperature of 5000\,K resides in a portion of temperature-radius space occupied by evolved red giants in the LMC, and we support this identification as well. We emphasize that neither of these two stars lie close enough to the center of the remnant to be a surviving companion to the exploded star, so we prefer these alternative explanations. We do not include either of the two stars discussed here in our Kernel Density Estimation that we use to compare to the population of surviving companion candidates due to their likely non-LMC main sequence identities.

\subsection{Comparison To Surviving Companion Models}

We have gathered a compilation of models that make concrete predictions about the effective temperature and radius of a surviving SN Ia companion on timescales appropriate for the SNR \snrem ($10^2 - 10^3$\,yrs), shown in Table \ref{tab:models}.  The considered surviving companion models shown predict that the star will either be hotter than $10000$\,K or inflated to larger than 2.5\,R$_\odot$. Our set of observations is sensitive to stars hotter than 5000\,K or larger than 0.6\,R$_\odot$. Therefore, we rule out this set of surviving companions from existing within SNR \snremspace, as each model  would be distinguishable from the sample of potential surviving companions in Figure \ref{fig:teff_rad}. 

Other surviving companion models not considered here can be easily tested and sought out in this remnant by comparing to our characterized stellar catalog. We show examples of our fit parameters in Table \ref{tab:fits}, and provide the full characterized catalog online. 

\begin{table*}[!ht]
    \centering
    \begin{tabular}{|l|l|l|l|l|l|l|l|l|l|l|l|}
    \hline
        MatchID & MatchRA & MatchDec & teff & teff\_err & rad & rad\_err & mh & mh\_err & av & av\_err & Near\_center \\ \hline
        1105108 & 77.37689 & -67.52006 & 5658.51 & 352.269 & 0.533 & 0.064 & -0.33 & 0.153 & 0.285 & 0.134 & False \\ \hline
        1240861 & 77.3773 & -67.52317 & 5414.569 & 244.166 & 0.54 & 0.036 & -0.349 & 0.145 & 0.269 & 0.137 & False \\ \hline
        2094417 & 77.37589 & -67.52166 & 6607.698 & 94.121 & 1.489 & 0.014 & -0.872 & 0.129 & 0.037 & 0.054 & False \\ \hline
        2360222 & 77.3794 & -67.52045 & 6650.82 & 54.514 & 1.381 & 0.012 & -0.424 & 0.136 & 0.027 & 0.03 & True \\ \hline
        4275360 & 77.38571 & -67.52132 & 5655.744 & 225.258 & 0.617 & 0.031 & -0.348 & 0.152 & 0.313 & 0.136 & False \\ \hline
        4304067 & 77.3827 & -67.52268 & 5203.318 & 193.617 & 0.716 & 0.035 & -0.352 & 0.146 & 0.311 & 0.138 & False \\ \hline
        4705369 & 77.37783 & -67.52192 & 5789.892 & 210.41 & 0.646 & 0.027 & -0.363 & 0.142 & 0.28 & 0.134 & True \\ \hline
        4728380 & 77.3771 & -67.51996 & 4997.528 & 8.752 & 3.896 & 0.013 & -0.96 & 0.057 & 0.003 & 0.004 & False \\ \hline
        4769912 & 77.37471 & -67.52032 & 5758.987 & 189.441 & 0.761 & 0.025 & -0.364 & 0.141 & 0.263 & 0.13 & False \\ \hline
        5394004 & 77.37623 & -67.51961 & 5161.428 & 16.844 & 2.683 & 0.015 & -0.721 & 0.114 & 0.007 & 0.01 & False \\ \hline
    \end{tabular}
    
    \caption{An abbreviated example of 10 stars with their extracted parameters. The full table including the magnitudes used for fitting is available as a supplemental online data product.}
    \label{tab:fits}

\end{table*}

\subsection{Results of Stars Identified in A22}
\label{subsec:faint_stars}
We additionally modeled the three stars identified in A22 that reside very near the dynamical center of SNR \snrem, within 1.4 arcsec corresponding to a velocity  $V \approx 1000$\,km\,s$^{-1}$ at the distance of the remnant. They are fainter than the detection limits of the HSC, so they are unsurprisingly the smallest and coolest stars in Figure \ref{fig:teff_rad}. All three stars appear to be a natural extension of the temperature-radius main sequence of the general LMC stellar population where M-dwarfs are expected to reside, and also lie within the contours estimated by the surrounding population, despite initial differences in aperture photometry. However, one of the three stars, which we refer to as HHE 5 in line with \cite{hovey_kinematic_2016}, is poorly fit by our models and we discuss it further in Appendix \ref{app:hhe5}. Like the other two stars, we support an ordinary M-dwarf classification for HHE 5.

\section{Discussion} \label{sec:discussion}

We have conducted a deep, systematic search of the Ia remnant SNR \snremspace to identify a surviving companion, using a sample of existing archival HST wide band photometry. We investigated a large enough search region to allow for the fastest moving surviving companions, i.e. hypervelocity white dwarfs tracing back to double detonation progenitor systems \citep[e.g.][]{shen_three_2018}. All models, regardless of the explosion mechanism of the primary, predict that a nearby secondary that is impacted by expanding supernova ejecta will show some combination of high effective temperature or inflated radius for at least thousands of years. In Table \ref{tab:models}, we show a compilation of expected lower temperatures and radii predicted from various models approximately 1000 years after being impacted by SN Ia ejecta. In contrast, none of the stars inside the SNR \snremspace close enough to the center to be surviving companion candidates in the remnant show astrophysical parameters that suggest exotic identities. For any of the models considered here, this finding rules out the existence of such a surviving companion as such a star would be clearly identifiable as separate from the main sequence stellar population. We note that we cannot rule out the existence of a He WD companion in line with \cite{shen_wait_2017} in SNR \snrem. However, this model solely considered luminosity generated due to the delayed decay of $^{56}$Ni on the surface of the donor WD, which may not remain the dominant source of over-luminosity at timescales associated with the remnant \citep[see e.g][]{pan_search_2014, liu_observational_2021}.

We rule out the existence of a surviving companion in the remnant down to a limiting optical AB magnitude of 25.5, with the one exception being that the surviving companion is indistinguishable from the local stellar population in combined temperature-radius space. The list of considered surviving companion models we show is not exhaustive, and new models not yet explored or published are still in development. We encourage models not considered here to be compared to the astrophysical properties of this stellar population. 

\subsection{Spin-up/spin-down models}

This work probes the interior stellar population of SNR \snremspace for  signatures of interaction both before and during the explosion of the primary WD. The models we show in Table \ref{tab:models} generally assume that the surviving companion is close enough to the primary when it explodes for the secondary to experience Roche lobe overflow. However, some models have been proposed in which the secondary donates mass to the primary C/O WD before the process halts and the secondary evolves in isolation. The mass transfer process imparts additional angular momentum onto the primary C/O WD which prevents it from experiencing thermonuclear runaway until it can dissipate the accrued angular momentum or spin down \citep{di_stefano_spin-upspin-down_2011, justham_single-degenerate_2011, hachisu_single_2012}. The spin-down timescale is poorly constrained to be between $10^5$ and $10^9$ years (but see \citealt{kerzendorf_search_2018} and enclosed references for further  constraints), and the secondary may have time to exhaust hydrogen evolve in to a WD. If so, the secondary can retain a large enough orbital separation from the primary to remain largely unaffected by the explosion. The secondary will then remain near the center of the remnant as an isolated WD that has cooled for up to the spin-down time. No isolated WD is detected in the archival photometry used in this work near the center of SNR \snremspace, which can be translated to a constraint on spin-down timescale of the system that is dependent on the mass of the companion (see \citealt{di_stefano_absence_2012}, but also \citealt{meng_constraining_2013} for further discussion).

\section{Conclusions} \label{sec:conclusions}

We present a full stellar classification of the interior stellar population to SNR \snremspace to probe the parameter space sensitive to ejecta interaction and search for a surviving companion to the exploded star that created the remnant. We considered a generous search region corresponding to an ejection velocity of 2500\,km\,s$^{-1}$ and a remnant expansion age of 400 years. Within this region, we do not detect an anomalously hot or radially extended star as predicted by the interaction of supernova ejecta with a nearby companion in both the degenerate and non-degenerate companion cases in which the companion survives. 

This result holds consistent with other recent non-detections of surviving companions in Ia remnants that may point to SN Ia progenitor channels that do not leave bound stellar remnants behind (see \citealt{ruiz-lapuente_surviving_2019}). However, the sporadic nature of previous surviving companion searches with inconsistent limits and parameter space exploration makes conclusive statements about the viability of specific progenitor channels difficult to support.  

The constraints of the LMC as well as the new developments to the HSC that made the stellar characterization central to this work possible remain consistent across other known SN Ia remnants in the LMC. Similar studies of those remnants would enable statistical statements about the viability of specific progenitor channels.

\section{Acknowledgments}

This work was based on observations made with the NASA/ESA Hubble Space Telescope, and obtained from the Hubble Legacy Archive, which is a collaboration between the Space Telescope Science Institute (STScI/NASA), the Space Telescope European Coordinating Facility (ST-ECF/ESAC/ESA) and the Canadian Astronomy Data Centre (CADC/NRC/CSA). PA and JPH acknowledge support from HST GO Program 16161 (PI: Arunachalam). JPH, the George A.\ and Margaret M.\ Downsbrough Professor of Astrophysics, acknowledges the Downsbrough heirs and the estate of George Atha Downsbrough for their support.

The data presented in this paper were obtained from the Mikulski Archive for Space Telescopes (MAST) at STScI. The specific observations analyzed can be accessed via \dataset[DOI: 10.17909/T97P46]{https://doi.org/10.17909/T97P46} through the HSC CasJobs database.

\software{\texttt{Astropy} \citep{the_astropy_collaboration_astropy_2013, astropy_collaboration_astropy_2018},
\texttt{Scipy} \citep{virtanen_scipy_2020},
\texttt{Numpy} \citep{harris_array_2020},
\texttt{pandas} \citep{reback_pandas-devpandas_2022},
\texttt{Matplotlib} \citep{hunter_matplotlib_2007},
\texttt{corner} \citep{foreman-mackey_cornerpy_2016},
\texttt{StarKit} \citep{kerzendorf_starkit_2015},
\texttt{wsynphot} \citep{kerzendorf_wsynphot_2022},
\texttt{APLpy} \citep{robitaille_aplpy_2012},
\texttt{UltraNest} \citep{buchner_ultranest_2021}.
}

\bibliography{references}{}
\bibliographystyle{aasjournal}



\section{Contributor Roles}

\begin{itemize}
\item{Conceptualization: Joshua V. Shields, John P. Hughes, Prasiddha Arunachalam}
\item{Data curation: Joshua V. Shields, Sofia Biriouk}
\item{Formal Analysis: Joshua V. Shields}
\item{Funding acquisition: Wolfgang Kerzendorf, John P. Hughes}
\item{Investigation: Joshua V. Shields, Prasiddha Arunachalam}
\item{Methodology: Joshua V. Shields}
\item{Project administration: Wolfgang Kerzendorf}
\item{Software: Johannes Buchner, Wolfgang Kerzendorf, Joshua V. Shields}
\item{Supervision: Wolfgang Kerzendorf, John P. Hughes, Joshua V. Shields} 
\item{Validation: Joshua V. Shields, Prasiddha Arunachalam, Sofia Biriouk}
\item{Visualization: Joshua V. Shields, Prasiddha Arunachalam, Hayden Monk, John P. Hughes}
\item{Writing - original draft: Joshua V. Shields}
\item{Writing - review \& editing: Joshua V. Shields, Wolfgang Kerzendorf, Prasiddha Arunachalam, John P. Hughes, Johannes Buchner}

\end{itemize}

\appendix \label{sec:appendix}

\section{Methodology verification - The Sun} \label{app:sun_val}

\begin{figure}[th]
    \centering
    \includegraphics[width=\textwidth]{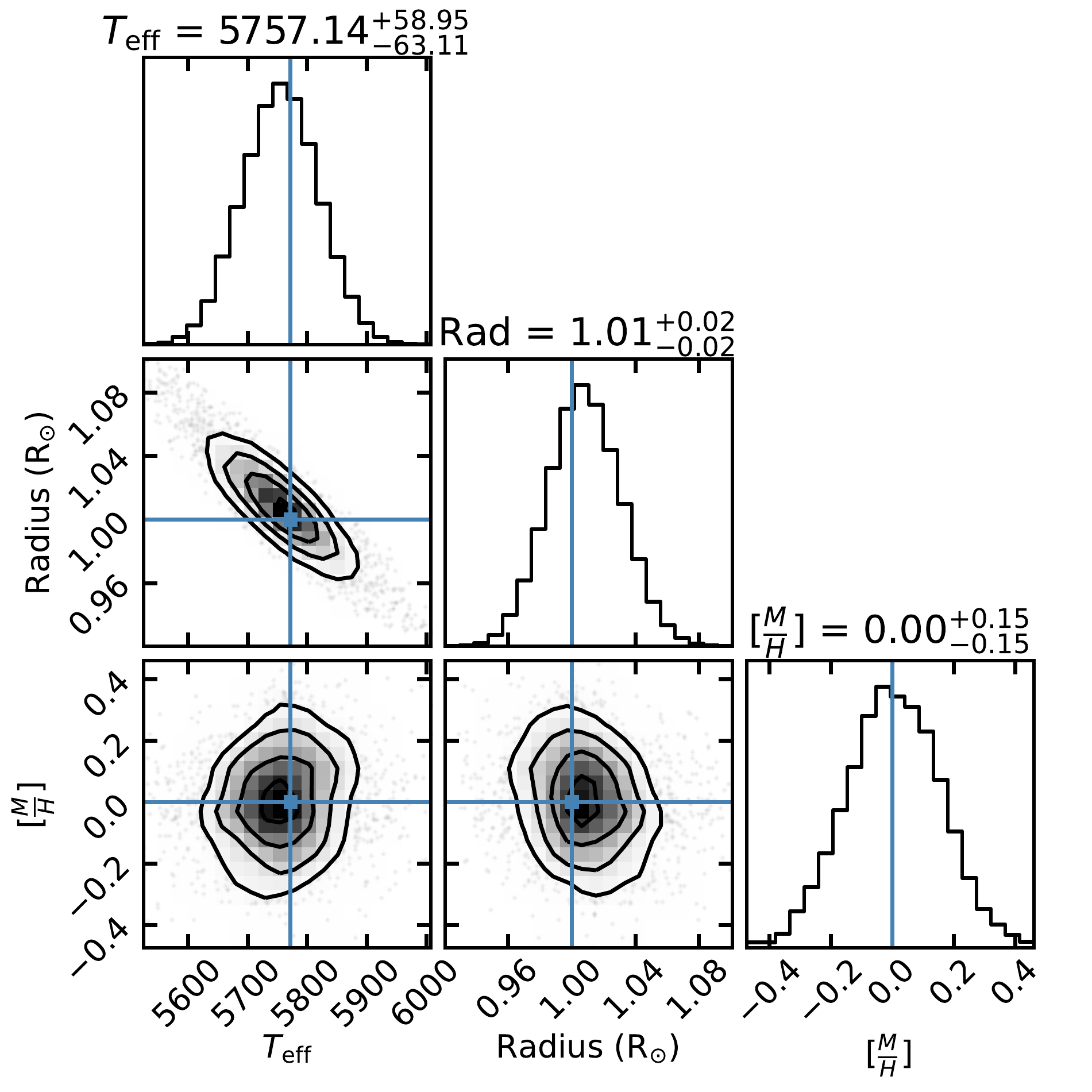}
    \caption{An example fit corner plot of extracted astrophysical parameters of the Sun for methodological validation. We obtained and fit magnitudes from \cite{willmer_absolute_2018}. The true values are shown with blue lines. Our parameter extraction shows strong agreement with the true solar values of temperature and radius, within $1\sigma$. 
}
    \label{fig:sun_compare}
\end{figure}

To validate our stellar astrophysical parameter extraction, we obtained AB magnitudes and magnitude uncertainties of the Sun from \cite{willmer_absolute_2018}. We applied the same methodology that we used in our study to these observations, excluding our treatment of dust, to recover derived solar astrophysical parameters. Specifically, we obtained the same five HST band AB magnitudes from our work, F475W, F555W, F875W, F110W, and F160W, and performed Bayesian inference to extract stellar parameters which we could then compare to known intrinsic values. The results of the parameter extraction are shown in Figure \ref{fig:sun_compare}. We used the same uninformative uniform temperature and radius priors as in the rest of our work but changed the metallicity prior to a normal distribution centered on solar metallicity (Mean = 0.00, $\sigma = 0.15$). As shown, our methodology powerfully constrains the temperature and radius of the star, producing strong agreement with true values, but once again recovering the prior on metallicity. See Appendix \ref{app:metallicity_val}
for further discussion on our metallicity constraints. 

\section{Measuring Metallicity With Photometry} \label{app:metallicity_val}

\begin{figure}[ht]
    \centering
    \includegraphics[width=\textwidth]{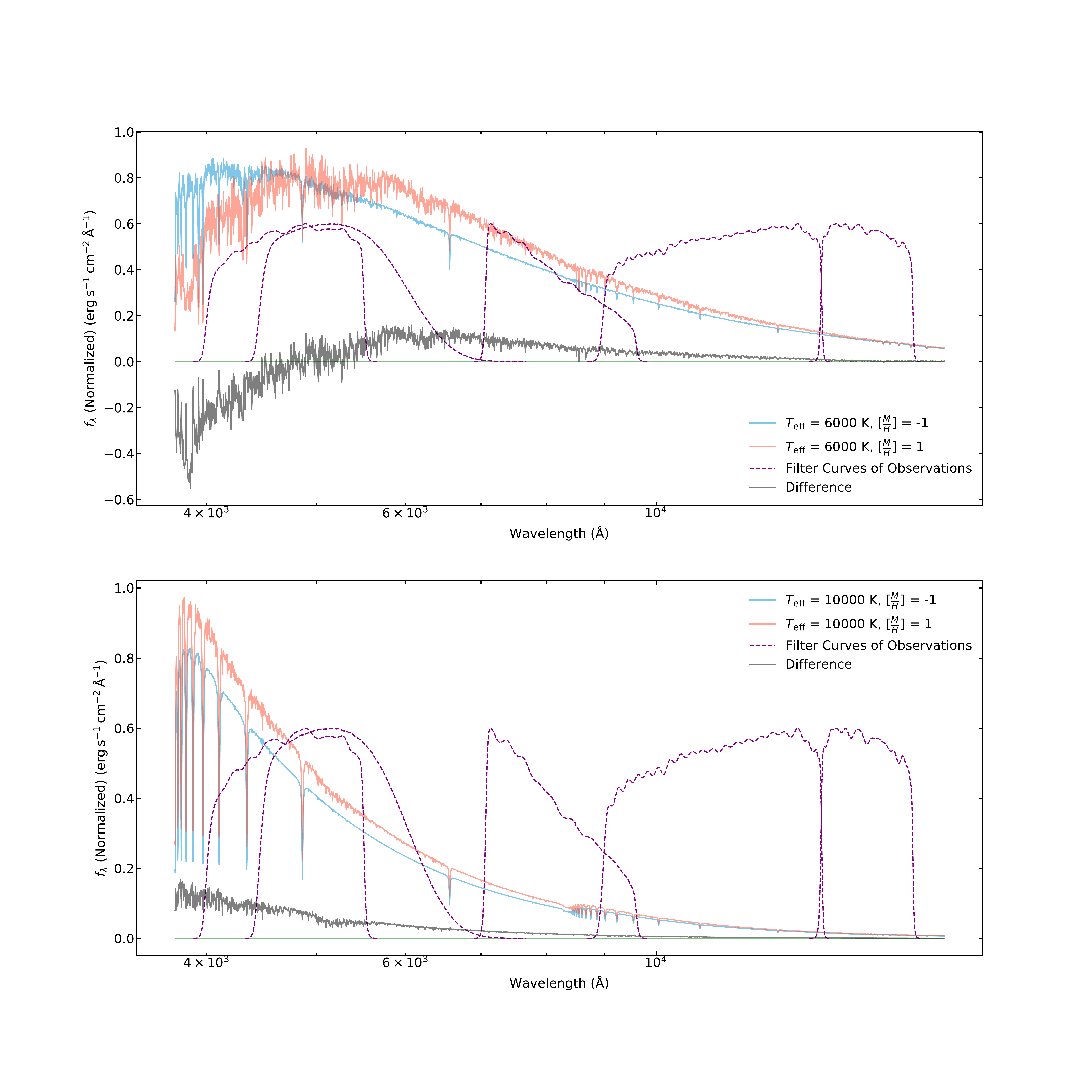}
    \caption{An example showing the effects of metallicity at different temperatures. The top panel shows the spectra of a metal-poor ($[\frac{M}{H}] = -1$) 6000\,K star and a metal-rich ($[\frac{M}{H}] = 1$) 6000\,K star. The bottom shows the same, but comparing 10000\,K stars. The purple dashed lines show the transmission curves of the same filters used in the rest of this work, and the gray line shows the difference between the two spectra. At some temperatures, sufficiently sensitive photometry is capable of constraining metallicity, detecting relative enhancements and diminishments in the redder and bluer optical portions of the spectrum. 
}
    \label{fig:metallicity_compare}
\end{figure}

We report metallicity measurements for each of the stars in this work but do not attempt to quantitatively probe this space for surviving companions. We made this choice because both the general ability of pure photometry and the combination of filters required to detect metallicity enhancement in a star has a complex relationship with effective temperature. In Figure \ref{fig:metallicity_compare} we show the relative differences in flux between metal-poor ($[\frac{M}{H}] = -1$)  and metal-rich ($[\frac{M}{H}] = 1$) stars at 6000\,K and 10000\,K. For the filters used in this work, [F475W, F555W, F814W, F110W, and F160W] these stars would show differences of [-0.04, 0.07, 0.17, 0.13, 0.04] mags and [0.15, 0.15, 0.14, 0.11, 0.11] mags at 6000\,K and 10000\,K respectively. The difference in metallicity for both temperature stars is possibly constrained by photometry, but this enrichment in a 6000\,K star creates a more obvious signature, distinguishable from a change in temperature or radius. 

In either case, the level of metallic enrichment expected for a surviving companion could potentially be constrained from the photometry of this work, as the differences in HST magnitudes from comparing the spectra in Figure \ref{fig:metallicity_compare} are significantly higher than the uncertainties of many observations, especially in the brighter stars in our sample. However, none of the stars in our sample showed an obvious signature tracing back to a departure from typical LMC metallicities, so the metallicities of stars in our sample are usually not constrained beyond our prior.


\section{Investigating HHE 5}
\label{app:hhe5}

\begin{figure}[ht]
    \centering
    \includegraphics[width=\textwidth]{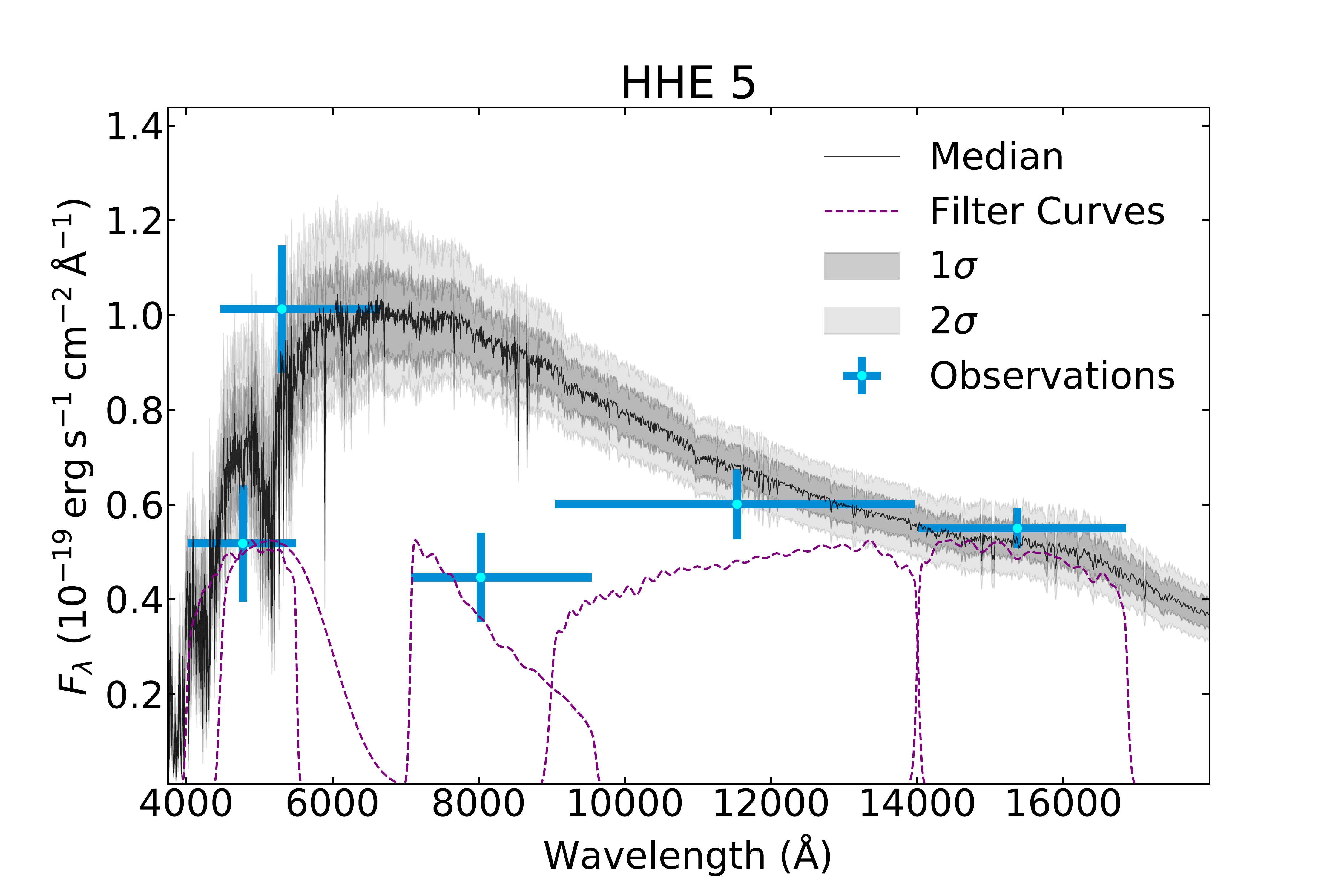}
    \caption{Our fit for HHE 5, similar to that shown in Figure \ref{fig:fit_example}. The model does a poor job fitting the observations, primarily because of the single F814W observation that is impossible to explain with a stellar spectrum capable of fitting the other four observations. The other points can be suitably explained by a typical stellar spectrum, in accordance with the possibility of an unphysical F814W measurement. 
}
    \label{fig:hhe5}
\end{figure}

In Figure \ref{fig:hhe5}, we show our fit of the SED belonging to HHE5 and originating from photometry in \cite{hovey_kinematic_2016}. The F814W measurement lies multiple standard deviations away from the constrained spectrum, and in general cannot lie on a standard stellar spectrum anchored by the infrared observations. Investigating further, we note that the F814W measurement of HHE 5 is the faintest F814W measurement present in the reported catalog of \cite{hovey_kinematic_2016}. While not conclusive, this could point to the measurement being unphysical. 

The location of HHE 5, very near the dynamical center of SNR \snrem, is the strongest argument in favor of a surviving companion identification, and we acknowledge that the unusual photometry here could be seen as further support towards an exotic identification. However, no current surviving companion models predict the narrow F814W suppression seen in the SED. We support the explanation that the F814W measurement is unphysical, rather than invoking an exotic scenario capable of explaining this feature.

\end{document}